# Massless Neutrino Oscillations via Quantum Tunneling

Hai-Long Zhao

**Abstract:** In the current theory, neutrino oscillations require non-vanishing masses of the neutrinos. By analogy with the oscillation of quantum two-state system, we assume that neutrino oscillations may be regarded as quantum tunneling process. The difference of the quantum numbers between two neutrinos may be regarded as a barrier between them. Thus neutrinos with vanishing mass can also oscillate. The hypothesis can also be applied to the oscillations of charged leptons, quarks, neutral mesons, as well as electroweak mixing.

## 1 Introduction

The experiments with solar, atmospheric and reactor neutrinos have provided compelling evidences for the existence of neutrino oscillations [1–11]. According to the current theory, neutrino oscillations are due to non-vanishing masses. Neutrino oscillations include oscillations in vacuum and in matter. The Seesaw mechanism has been proposed to explain why the neutrino masses are so small [12, 13]. The oscillation length can be derived by the current theory [14]. But some of the assumptions, such as equal-energy or equal-momentum, are controversial. This has led to the wave packet description of neutrinos [15–25]. However, both the two descriptions face a problem that the mixing will become incoherent and neutrinos will cease oscillation after a long distance of flight.

In order to realize neutrino oscillations within the framework of standard model, we assume that neutrino oscillations are analogous to the oscillation of quantum two-state system, then massless neutrinos can also oscillate.

## 2 The current theory of neutrino oscillations

As a plane wave, neutrino can oscillate via four assumptions: equal-energy, equal-momentum, energy-momentum conservation and equal-velocity. For a review of these assumptions one may see [24, 25]. We mainly discuss the former two assumptions, which can be found in [14]. For the latter two assumptions, one may see [24, 25] and the references therein.

For convenience we work in the natural units, where $\hbar = c = 1$. For simplicity, we leave the tauon neutrino $v_\tau$ out of the following and assume that only the electron neutrino $v_e$ and muon neutrino $v_\mu$ mix with each other. As $v_e$ and $v_\mu$ are not the energy eigenstates, we denote the eigenstates of the Hamiltonian with $v_1$ and $v_2$, respectively, and for which we make the following ansatz

$$\begin{pmatrix} v_1 \\ v_2 \end{pmatrix} = \begin{pmatrix} \cos\theta & -\sin\theta \\ \sin\theta & \cos\theta \end{pmatrix} \begin{pmatrix} v_e \\ v_\mu \end{pmatrix}. \tag{1}$$

The inversion of Eq. (1) is

Email: zhlzyj@126.com

$$\begin{pmatrix} \nu_e \\ \nu_\mu \end{pmatrix} = \begin{pmatrix} \cos\theta & \sin\theta \\ -\sin\theta & \cos\theta \end{pmatrix} \begin{pmatrix} \nu_1 \\ \nu_2 \end{pmatrix}. \tag{2}$$

First we assume that an electron neutrino $\nu_e$ is produced with definite momentum $\mathbf{p}$ at point $\mathbf{x}$ and at time $t=0$. In the energy representation it holds that, for the evolution of the state

$$\begin{pmatrix} \nu_1(\mathbf{x},t) \\ \nu_2(\mathbf{x},t) \end{pmatrix} = \begin{pmatrix} e^{-iE_1 t} & 0 \\ 0 & e^{-iE_2 t} \end{pmatrix} \begin{pmatrix} \nu_1(0) \\ \nu_2(0) \end{pmatrix} e^{i\mathbf{p}\cdot\mathbf{x}}, \tag{3}$$

where $E_1$ and $E_2$ are the energy eigenvalues of $\nu_1$ and $\nu_2$, respectively, and we have

$$E_1 = \sqrt{p^2 + m_1^2}\,, \qquad E_2 = \sqrt{p^2 + m_2^2}\,, \tag{4}$$

where $m_1$ and $m_2$ are the masses of the eigenstates $\nu_1$ and $\nu_2$, respectively. With the help of above equations, we obtain

$$\begin{pmatrix} \nu_e(\mathbf{x},t) \\ \nu_\mu(\mathbf{x},t) \end{pmatrix} = \begin{pmatrix} \cos\theta & \sin\theta \\ -\sin\theta & \cos\theta \end{pmatrix} \begin{pmatrix} e^{-iE_1 t} & 0 \\ 0 & e^{-iE_2 t} \end{pmatrix} \begin{pmatrix} \cos\theta & -\sin\theta \\ \sin\theta & \cos\theta \end{pmatrix} \begin{pmatrix} \nu_e(0) \\ \nu_\mu(0) \end{pmatrix} e^{i\mathbf{p}\cdot\mathbf{x}}$$

$$= \begin{pmatrix} \cos^2\theta e^{-iE_1 t} + \sin^2\theta e^{-iE_2 t} & \sin\theta\cos\theta(e^{-iE_2 t} - e^{-iE_1 t}) \\ \sin\theta\cos\theta(e^{-iE_2 t} - e^{-iE_1 t}) & \cos^2\theta e^{-iE_1 t} + \sin^2\theta e^{-iE_2 t} \end{pmatrix} \begin{pmatrix} \nu_e(0) \\ \nu_\mu(0) \end{pmatrix} e^{i\mathbf{p}\cdot\mathbf{x}}. \tag{5}$$

According to the assumption that pure $\nu_e$ are emitted at the source, we have $\nu_e(0)=1$ and $\nu_u(0)=0$. Then the probability of finding $\nu_u$ at time $t$ is

$$|\nu_\mu(\mathbf{x},t)|^2 = |\sin\theta\cos\theta(e^{-iE_2 t} - e^{-iE_1 t})|^2 = \sin^2(2\theta)\sin^2\frac{(E_2 - E_1)t}{2}. \tag{6}$$

If $m_1 = m_2 = 0$, we have $E_1 = E_2$, then the above expression is equal to zero and neutrinos cannot oscillate. In relativistic limit, we have

$$E_2 - E_1 = \sqrt{m_2^2 + p^2} - \sqrt{m_1^2 + p^2} \approx \frac{(m_2^2 - m_1^2)}{2p} \tag{7}$$

Let $\Delta m_{21}^2 = (m_2^2 - m_1^2)$ and $t \approx l$, the oscillation length $l$ can be written as

$$l = \frac{4\pi p}{\Delta m_{21}^2} \tag{8}$$

We then suppose that neutrinos are produced with a definite energy $E$, and make the ansatz

$$\begin{pmatrix} \nu_1(\mathbf{x},t) \\ \nu_2(\mathbf{x},t) \end{pmatrix} = \begin{pmatrix} e^{ip_1 x}\nu_1(0) \\ e^{ip_2 x}\nu_2(0) \end{pmatrix}, \tag{9}$$

where $p_1 = \sqrt{E^2 - m_1^2}$, $p_2 = \sqrt{E^2 - m_2^2}$. And again set $\nu_e(0)=1$ and $\nu_u(0)=0$ for $x=0$. It results that

$$|\nu_\mu(\mathbf{x},t)|^2 = \sin^2(2\theta)\sin^2\frac{(p_2 - p_1)x}{2}. \tag{10}$$

In relativistic limit, we have

$$p_2 - p_1 = \sqrt{E^2 - m_2^2} - \sqrt{E^2 - m_1^2} \approx \frac{(m_2^2 - m_1^2)}{2E} \tag{11}$$

When the neutrinos fly at nearly the speed of light, we have $x \approx t$ and $E \approx p$. Then Eq. (11) is identical to Eq. (6). Besides the assumptions of equal–momentum and equal–energy, there are also assumptions of energy–momentum conservation and equal–velocity. Although the four assumptions lead to the same result in the relativistic limit, no one gives arguments for their correctness. A first look at the four assumptions shows that they are incompatible. For example, if the neutrinos with different masses had the same energy they could not have the same momentum and vice versa. The assumption of energy–momentum conservation seems to be the most satisfying one, while the equal–velocity assumption is the most unlikely one. An explanation in [26] was proposed to account for why equal–velocity assumption can be ruled out. Assuming the two mass eigenstates have a same velocity $u$, we immediately arrive at $\gamma_1 = \gamma_2 = \sqrt{1-u^2}$, then we get $E_1 / E_2 = m_1 / m_2$. This equality cannot hold because $E_1 / E_2 \approx 1$, while $m_1 / m_2$ may be extremely small or extremely large.

Since the plane wave description of neutrino is unsatisfactory, it's natural to describe it with wave packet. The detailed discussion may refer to [15–25]. But a problem remains for the two descriptions, that is, since the equal–velocity assumption is not correct, different neutrinos will travel with a different velocity. Then after a long distance of flight (e.g. the neutrinos reaching the earth from supernovas), the neutrino mixing will become incoherent and neutrinos will cease oscillation. This difficulty will be swept away with massless neutrino oscillations model in the following.

The above analysis of two–flavor neutrino mixing can be easily extended to the instance of three–flavor mixing, which is described with PMNS matrix [27].

## 3 Neutrino oscillations via barrier tunneling

### 3.1 Oscillation of quantum two-state system

There is a two-state system in quantum physics, which has two quantum states with a symmetric structure. The system will oscillate between the two states. The following discussion may refer to [28]. Taking ammonia molecule as an example. There are two possible positions for the nitrogen atom, which may be on one side of the plane or on the other, as shown in Fig. 1. We denote the two quantum states by $|C_1\rangle$ and $|C_2\rangle$, respectively. In Fig. 1, the nitrogen atom must penetrate a barrier when flipping to the other side. Even if its energy is not high enough to traverse the barrier from the classical point of view, there is a certain probability for the nitrogen atom to tunnel through the barrier. So we suppose the Schrödinger equation for the flipping of the nitrogen atom to be

$$i\hbar\frac{\partial}{\partial t}\begin{pmatrix} C_1 \\ C_2 \end{pmatrix} = \begin{pmatrix} E_0 & -A \\ -A & E_0 \end{pmatrix}\begin{pmatrix} C_1 \\ C_2 \end{pmatrix}. \tag{12}$$

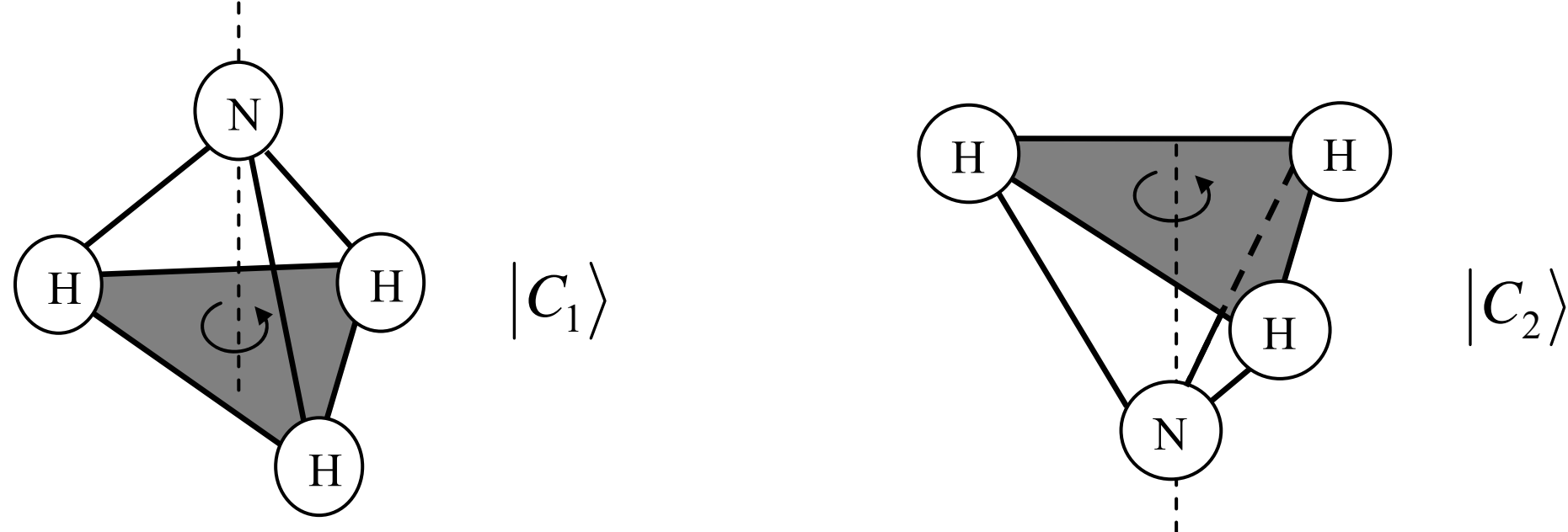

Fig. 1 Two symmetric configurations of ammonia molecule.

It's not difficult to solve this differential equation. The detailed calculations may refer to [28]. We only give the result here. Suppose the system is in state $|C_1\rangle$ at $t=0$, i.e., $C_1(0)=1$, $C_2(0)=0$. Then the probabilities of the system in states $|C_1\rangle$ and $|C_2\rangle$ are

$$\begin{cases} C_1(t) = e^{-iE_0 t/\hbar} \dfrac{\cos At}{\hbar} \\ C_2(t) = e^{-iE_0 t/\hbar} \dfrac{\sin At}{\hbar} \end{cases}. \tag{13}$$

respectively. The eigenvalues of the energy of the system are $E_1 = E_0 - A$ and $E_2 = E_0 + A$, respectively. The system will oscillate between $|C_1\rangle$ and $|C_2\rangle$.

### 3.2 Neutrino oscillations via barrier tunneling

We see from Eq. (13) that the system will oscillate between $|C_1\rangle$ and $|C_2\rangle$, and Eqs. (6) and (13) are identical in the case of $\theta = \pi/4$ and $A = (E_2 - E_1)/2$. So if we wish to let neutrinos oscillate with vanishing mass, we may simply think that neutrino oscillations are the consequence of barrier tunneling. In this case, even if $m_1 = m_2 = 0$, we still have $E_1 \neq E_2$, where $E_1$ and $E_2$ are the eigenvalues of the energy of the system. The situation of Eq. (13) is complete oscillation, that is, the system will be in a pure state ($|C_1\rangle$ or $|C_2\rangle$) periodically. The more general situation is that only part of the two quantum states are involved in the oscillation, then a mixing angle $\theta$ will appear, as shown in Eq. (1), which describes the mixing of the two quantum states. It should be noted that in Eq. (7), the difference between $E_2$ and $E_1$ comes from the mass difference of the neutrinos; while in the barrier tunneling theory, the energy difference of the system comes from the potential energy of the barrier.

A problem with quantum barrier tunneling is that we only know that the difference of the eigenvalues of energy of the system is due to the existence of barrier, but we can not determine the expression of $A$. We need to find out the relationship between $A$ and the energy of the neutrino. Because the energy splitting is very small, we may use the average energy $E$ to denote $E_1$ or $E_2$. Since there is no prior guideline to follow, we may use Bohr's correspondence principle to seek some clues. This principle states that quantum mechanics reduces to classical mechanics in the limit of large quantum numbers, while large quantum numbers implies high system energy, therefore when the neutrino energy is very high, the neutrino is more like a classical particle that is in the superposition of two states rather than oscillation between them, that is, the energy splitting will tend to decreases with the increase of neutrino energy. Then we may assume that

$$A = \frac{1}{2}\Delta E = \frac{k}{E^n} \tag{14}$$

From the fitting of experimental results we have $n \approx 1.14$ [29], so a reasonable assumption is $n = 1$. Then we get the same result as Eq. (8), and $k$ is equivalent to $\Delta m_{21}^2$ in Eq. (8).

We have pointed out earlier that the assumption of equal-energy or equal-momentum is controversial when using the plane wave model. This difficulty is overcome when quantum tunneling hypothesis is adopted. In our theory, the neutrino mass is zero, and the energies of the three flavor neutrinos are the same, but due to the existence of barriers, the eigenvalues of energy of the system are different, so the system can still oscillate. Our model can also explain the phenomenon of neutrino oscillation enhancement in matter. Since only $v_e$ can interact with matter, which is equivalent to that there is a potential energy in addition to the kinetic energy when $v_e$ passes through matter and its total energy increases, it will be easier for $v_e$ to flip to $v_\mu$ or $v_\tau$.

There are also other massless neutrino oscillations models. But they all need some additional assumptions. For example, residual symmetry in [30], open system in [31] and modified Dirac equations in [32], are respectively introduced in order to keep neutrinos massless. By contrast, our model is the simplest, and it is natural to extend the oscillation of quantum two-state system to neutrino oscillations.

### 3.3 Formation mechanism of barrier

We have adopted the hypothesis of barrier tunneling to explain the neutrino oscillations, then the next question is: How are the barriers between neutrinos formed? Let's first see some tunneling phenomena. In the scanning tunneling microscope, electrons tunnel through air from one metal to another metal, the barrier is the air between the two metals. In the Josephson junction, the barrier is the insulator sandwiched between the superconductors. In the photon tunneling experiment, the barrier is the air between the two prisms [33, 34]. These tunneling phenomena are quantum effects, and they can be explained with quantum mechanics without introducing new interaction. The only requirement for quantum tunneling is the conservation of energy and momentum of the particles before and after tunneling.

Now let's see the electron transition outside the nucleus between different orbitals, which may also be regarded as a quantum tunneling phenomenon. The barrier here is not a tangible substance, but the difference in quantum numbers ($n$, $l$, $m$) of the electron in different orbitals. Due to the different energies of the electron in different orbitals, the electron will emit or absorb a photon during the transition process. As for the neutrinos, the difference in quantum number is the flavor violation between two generations of neutrinos. Therefore, we see that the different quantum numbers will build a barrier between two quantum states. When a neutrino makes a transition from $v_e$ to $v_\mu$ (or $v_\tau$), it must borrow the energy of $\Delta E$ from the vacuum, and then return the energy after the transition.

What if there are no barriers between different flavors of neutrino? In this case, neutrino cannot maintain the independence of flavor, it will become a simple mixture of three flavors. Its property will be different from any flavor of the neutrino. To illustrate this point, we take the polarization of light as an example. The linear polarization can be considered as the superposition of the left-handed and right-handed circular polarizations. There is no barrier between the two eigenstates. Thus the property of the linearly polarized light is different from that of the circularly polarized light. We see that the existence of the barrier of neutrino not only is the theoretical requirement but also has physical significance.

Let's consider another question: Why do neutrinos oscillate? We know that there are three generations of quarks in nature, and we can distinguish different quarks in terms of their masses, and so do the three generations of charged leptons. As for the neutrinos, if all the three generations of neutrinos have the same vanishing mass, how can we distinguish them? On the other hand, both theory and experiment require the existence of three generations of neutrinos. Therefore, neutrinos must oscillate.

We might also understand neutrino oscillations from the properties of microscopic particles. Microscopic particles are never satisfied with staying at the same place. They will appear in every place in the coherent volume simultaneously. Similarly, a microscopic particle is not satisfied to appear as only character. If possible, it will change its character from one to another, which is the intrinsic cause of neutrino oscillations. As flavor oscillation is an intrinsic property of fundamental particles, not only neutrinos can oscillate, but also other particles will oscillate. We shall see some examples in the following.

## 4 Oscillations of other particles

### 4.1 Oscillations of charged leptons

Similar to neutrino oscillations, three generations of charged leptons can also oscillate. However, due to the restriction of the conservation of energy and momentum, two particles with different masses cannot transform into each other freely. This can be seen from the following formulas:

$$m_1 v_1 = m_2 v_2 \tag{15}$$

$$\frac{m_1 v_1}{\sqrt{1 - v_1^2 / c^2}} = \frac{m_2 v_2}{\sqrt{1 - v_2^2 / c^2}} \tag{16}$$

If Eq. (15) holds, we get $v_1 = v_2$ from Eq. (16). Then Eq. (15) fails. So a free muon cannot directly transform into an electron, but it can do so by emitting a photon, that is, the following process may occur:

$$\mu^- \to e^- + \gamma \tag{17}$$

This is actually electromagnetic decay. However, this process is implemented by barrier tunneling rather than electromagnetic interaction. This process is analogues to the transition of electron between different orbitals. As the mass difference between electron and muon is very large, this probability is very small. Furthermore, this decay mode will be greatly suppressed by the muon weak decay, as the following process are more likely to occur:

$$\mu^- \to e^- + \nu_\mu + \overline{\nu}_e \tag{18}$$

The above reaction has not been observed experimentally so far. As the photon emitted in the reaction has a very large energy, it may also transformed into a positron and an electron, so two electrons and a positron will be observed in the experiment. The aim of the Mu3e experiment being planned is to observe the above result. It is worth pointing out that in order to implement the reaction of Eq. (17), an undiscovered particle X is needed in the current theory. The mass of X particle is of the order of $10^3$ TeV. The Feynman diagram of the reaction is shown in Fig. 2. According to our theory, the reaction of Eq. (17) can be realized by quantum tunneling without introducing X particle.

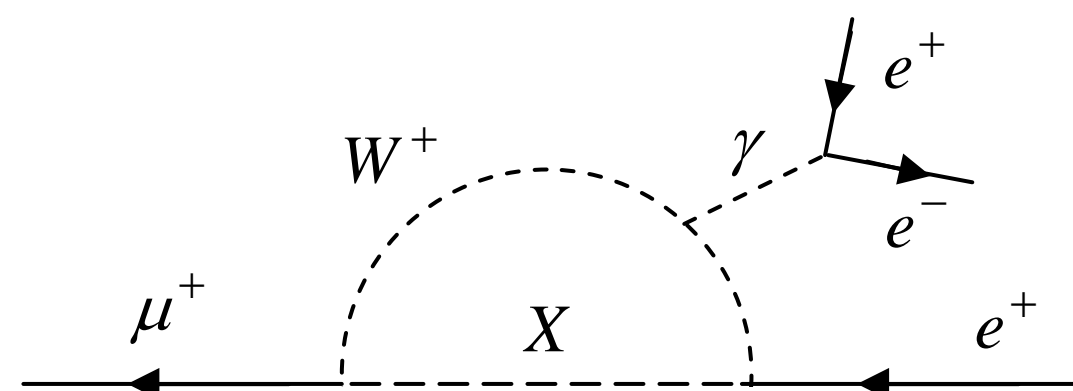

Fig. 2 Feynman diagram for Mu3e experiment.

### 4.2 Quark oscillations

The lepton oscillations can be extended to the situation of quarks. Since there are no free quarks, quark oscillations can only occur inside hadrons. Quarks in hadrons interact with other quarks at all times, that is, they also have potential energy. With the help of potential energy, quarks in hadrons may oscillate without destroying the conservation of energy and momentum. Similar to the situation of leptons, quark oscillations are also suppressed by weak decay, thus the oscillation probability is very small and can hardly be observed. Nevertheless, the experiments at the CERN Large Hadron Collider provide an opportunity to observe this event. The LHCb cooperation found a discrepancy of 4.1 σ between the decay probability of $B^0 \to K^{*0} \to K^+\pi^-\mu^+\mu^-$ and the expectation of the standard model [35]. In the decay process a b quark in the $B^0$ meson becomes an s quark. The standard model only considers the decay caused by weak interaction. If quark oscillation, i.e., the direct transition from b quark to s quark, is taken into account, we should be able to explain this discrepancy without introducing new particle beyond the standard model.

### 4.3 Oscillation of neutral kaons

We now see another oscillation phenomenon——the oscillation of neutral kaons. In the oscillation process, the CP quantum number of the particle is violated. The oscillation between $\mathrm{K}^0(\mathrm{d}\bar{\mathrm{s}})$ and $\bar{\mathrm{K}}^0(\bar{\mathrm{d}}\mathrm{s})$ is realized by weak interaction [14]. This process can be described as follows. The d quark in the $\mathrm{K}^0$ meson captures a virtual $\bar{\mathrm{u}}$ quark in the vacuum and becomes a $\mathrm{W}^-$ boson, which then decays into s and $\bar{\mathrm{u}}$ quarks; in the meanwhile, the $\bar{\mathrm{s}}$ in $\bar{\mathrm{K}}^0$ captures a virtual u quark in the vacuum and becomes a $\mathrm{W}^+$ boson, which then decays into $\bar{\mathrm{d}}$ and u quarks. U quark and $\bar{\mathrm{u}}$ quark annihilate into vacuum, and $\bar{\mathrm{d}}\mathrm{s}$ is left. By this way, a $\mathrm{K}^0$ meson will transform into a $\bar{\mathrm{K}}^0$ meson. The process can be written as

$$\mathrm{K}^0 \to \left\{ \begin{array}{l} \mathrm{d}+\bar{\mathrm{u}} \to \mathrm{W}^- \to \mathrm{s}+\bar{\mathrm{u}} \\ \bar{\mathrm{s}}+\mathrm{u} \to \mathrm{W}^+ \to \bar{\mathrm{d}}+\mathrm{u} \end{array} \right\} \to \bar{\mathrm{K}}^0 \tag{19}$$

With $\mathrm{K}^0$ and $\bar{\mathrm{K}}^0$ we can construct other two quantum states, namely

$$\mathrm{K}_1 = \frac{1}{\sqrt{2}}(\mathrm{K}^0 + \bar{\mathrm{K}}^0), \qquad \mathrm{K}_2 = \frac{1}{\sqrt{2}}(\mathrm{K}^0 - \bar{\mathrm{K}}^0) \tag{20}$$

As $\mathrm{K}_1$ and $\mathrm{K}_2$ have the same mass but different CP quantum numbers, i.e., $\mathrm{CP}\,\mathrm{K}_1 = \mathrm{K}_1$ and $\mathrm{CP}\,\mathrm{K}_2 = -\mathrm{K}_2$, there will be an oscillation between $\mathrm{K}_1$ and $\mathrm{K}_2$. The consequence of the oscillation is that neither $\mathrm{K}_1$ nor $\mathrm{K}_2$ is the eigenstate of the energy. What we observe experimentally are the mixture of $\mathrm{K}_1$ and $\mathrm{K}_2$, i.e., $\mathrm{K}_S^0 = \mathrm{K}_1 + \varepsilon_1 \mathrm{K}_2$ and $\mathrm{K}_L^0 = \mathrm{K}_2 + \varepsilon_2 \mathrm{K}_1$, where $\varepsilon_{1,2} \approx 2.2 \times 10^{-3}$. $\mathrm{K}_S^0$ ($0.89 \times 10^{-10}$ s) and $\mathrm{K}_L^0$ ($5.1 \times 10^{-8}$ s) are the eigenstates of energy, and there is a small energy difference of $\Delta mc^2 = 0.529 \times 10^{10}\, \hbar / s$ between them [36].

The oscillation between $\mathrm{K}^0$ and $\bar{\mathrm{K}}^0$ is realized by virtual particles, while the oscillation between $\mathrm{K}_1$ and $\mathrm{K}_2$ is realized by barrier tunneling. It can be vividly imagined that there is a bridge between $\mathrm{K}^0$ and $\bar{\mathrm{K}}^0$; while there is no bridge between $\mathrm{K}_1$ and $\mathrm{K}_2$, and the gap can be crossed by one large step.

When neutral kaons are born, half of them are in $\mathrm{K}_1$ state, and half of them are in $\mathrm{K}_2$ state. As the CP number of two pions is 1, while the CP number of three pions is -1, according to CP conservation, $\mathrm{K}_1$ will decay into two pions, while $\mathrm{K}_2$ will decay into three pions. Due to the oscillation between $\mathrm{K}_1$ and $\mathrm{K}_2$, a small fraction of $\mathrm{K}_2$ (about two thousandths) will first transform into $\mathrm{K}_1$ before decaying into two pions, resulting in CP violation.

It should be pointed out that the current theory attributes the phenomenon of $\mathrm{K}_L^0$ decaying into two pions to the CP violation, and does not explain how this process occurs. Our theory divides this process into two stages: The first one is $\mathrm{K}_2$ oscillating to $\mathrm{K}_1$, the second one is $\mathrm{K}_1$ decaying into two pions. Therefore, CP non-conservation of neutral kaons can be attributed to the oscillation between $\mathrm{K}_1$ and $\mathrm{K}_2$; while CP is still conserved during the decay process.

### 4.4 Oscillations of other neutral mesons

In the following, we shall see two other examples of oscillation of neutral mesons. First, let's see the oscillation of $\eta$ meson. It has two quantum states with the same energy, namely

$$\eta_8 = \frac{1}{\sqrt{6}}(u\bar{u} + d\bar{d} - 2s\bar{s}) , \qquad \eta_0 = \frac{1}{\sqrt{3}}(u\bar{u} + d\bar{d} + s\bar{s}) \tag{21}$$

where $\eta_8$ is octet state and $\eta_0$ the singlet. The difference in flavor builds a barrier, so there is oscillation between them. The $\eta$ and $\eta'$ particles observed experimentally are the mixture of $\eta_8$ and $\eta_0$, i.e.,

$$\eta = \eta_8 \cos\theta - \eta_0 \sin\theta , \qquad \eta' = \eta_8 \sin\theta + \eta_0 \cos\theta \tag{22}$$

where the mixing angle is $\theta \approx -10^\circ$ [37]. A similar example is the mixing of $\phi$ particle, which has two quantum states with the same energy:

$$\phi_8 = \frac{1}{\sqrt{6}}(u\bar{u} + d\bar{d} - 2s\bar{s}) , \qquad \phi_0 = \frac{1}{\sqrt{3}}(u\bar{u} + d\bar{d} + s\bar{s}) \tag{23}$$

The $\omega$ and $\phi$ particles observed experimentally are the mixture of $\phi_8$ and $\phi_0$, i.e.,

$$\omega = \phi_8 \sin\theta + \phi_0 \cos\theta , \qquad \phi = \phi_8 \cos\theta - \phi_0 \sin\theta \tag{24}$$

where the mixing angle is $\theta \approx 35^\circ$ [37].

### 4.5 Electroweak mixing

Beyond the electroweak unified energy scale, the properties of photon $\gamma$ and $Z^0$ boson are almost identical: They are electrically neutral; they have vanishing mass and 1 spin. On the other hand, there are differences between them. For example, photons can only interact with charged particles, while $Z^0$ boson can interact with both charged and neutral particles. This difference will build a barrier between photon and $Z^0$ boson, and there should exist oscillation between them. However, what we observed experimentally are $\gamma$ and $Z^0$, so the actual particles that oscillate are $W^0$ and B particles, $\gamma$ and $Z^0$ are the mixture of $W^0$ and B particles, and we have

$$A_\mu = B_\mu \cos\theta_W + W_\mu^0 \sin\theta_W , \qquad Z_\mu = -B_\mu \sin\theta_W + W_\mu^0 \cos\theta_W \tag{25}$$

where the mixing angle is Weinberg angle $\theta_W$, whose current experimental value is $\sin^2\theta_W = 0.2232$ [36].

## 5 Conclusion

Neutrino oscillations with vanishing mass lead to a simple version of standard model, and we have already had an appropriate equation to describe the behavior of neutrinos, i.e., Weyl equation. Barrier tunneling provides a simple explanation for neutrino oscillations, and the difficulty with the plane wave and wave packet descriptions will not exist. The barrier tunneling can be easily extended to quark mixing, forbidden transition of charged leptons, neutral meson oscillation and electroweak mixing. Such a hypothesis can explain a number of phenomena, indicating that it is reasonable.

As long as the neutrino mass is zero and there are three generations of neutrinos, neutrino is bound to oscillate, which is determined by the property of the microscopic particle itself. If a microscopic particle has multiple states and the energies of these states are the same, then the particle will oscillate between different states. It will never be satisfied with only existing in one state, it will try all possible states.

## References

[1] Y. Fukuda, et al., [Super-Kamiokande Collab.], Evidence for oscillation of atmospheric neutrinos, Phys. Rev. Lett. **81** (1998) 1562-1567.

[2] Q. R. Ahmad, et al., [SNO Collab.], Measurement of the rate of $v_e + d \to p + p + e^-$ interactions Produced by $^8B$ solar neutrinos at the Sudbury Neutrino Observatory, Phys. Rev. Lett. 87 (2001) 071301.

[3] Q. R. Ahmad, et al., [SNO Collab.], Direct evidence for neutrino flavor transformation from neutral-current interactions in the Sudbury Neutrino Observatory, Phys. Rev. Lett. 89 (2002) 011301.

[4] W. Hampel, et al., [GALLEX Collab.], GALLEX solar neutrino observations: Results for GALLEX IV, Phys. Lett. B 447 (1999) 127-133.

[5] M. Altmann, et al., [GNO Collab.], Complete results for five years of GNO solar neutrino observations, Phys. Lett. B **616** (2005) 174-190.

[6] K. Eguchi, et al., [KamLAND Collab.], First results from KamLAND: Evidence for reactor antineutrino disappearance, Phys. Rev. Lett. **90** (2003) 021802.

[7] M. H. Ahn, et al., [K2K Collab.], Measurement of neutrino oscillation by the K2K experiment, Phys. Rev. D **74** (2006) 072003.

[8] D. G. Michael, et al., [MINOS Collab.], Observation of muon neutrino disappearance with the MINOS detectors and the NuMI neutrino beam, Phys. Rev. Lett. **97** (2006) 191801.

[9] P. Adamson, et al., [MINOS Collab.], Measurement of neutrino oscillations with the MINOS detectors in the NuMI beam, Phys. Rev. Lett. **101** (2008) 131802.

[10] F. P. An, et al., [Daya Bay Collab.], Observation of electron-antineutrino disappearance at Daya bay, Phys. Rev. Lett. **108** (2012) 171803.

[11] J. K. Ahn, et al., [RENO Collab.], Observation of reactor electron antineutrinos disappearance in the RENO experiment, Phys. Rev. Lett. **108** (2012) 191802.

[12] T. Yanagida, Horizontal symmetry and masses of neutrinos, Prog.Theor. Phys. **64** (1980) 1103-1105.

[13] M. Lindner, T. Ohlsson and G. Seidl, Seesaw mechanisms for Dirac and Majorana neutrino masses, Phys. Rev. D **65** (2002) 053014.

[14] G. Greiner and B. Müller, Gauge theory of weak interactions. 3rd. ed., Springer-Verlag, 2000.

[15] S. Nussinov, Solar neutrinos and neutrino mixing, Phys. Lett. B **63** (1976) 201-203.

[16] B. Kayser, On the quantum mechanics of neutrino oscillation, Phys. Rev. D **24** (1981) 110-116.

[17] C. Giunti, C. W. Kim and U. W. Lee, When do neutrinos really oscillate?: Quantum mechanics of neutrino oscillations, Phys. Rev. D **44** (1991) 3635-3640.

[18] C. Giunti and C. W. Kim, Fundamentals of neutrino physics and astrophysics, Oxford University Press, 2007.

[19] C. Giunti, C. W. Kim, J. A. Lee and U. W. Lee, On the treatment of neutrino oscillations without

resort to weak eigenstates, Phys. Rev. D **48** (1993) 4310-4317.

[20] K. Kiers and N. Weiss, Neutrino oscillations in a model with a source and detector, Phys. Rev. D **57** (1998) 3091-3105.

[21] C. Y. Cardall, Coherence of neutrino flavor mixing in quantum field theory, Phys. Rev. D **61** (2000) 073006.

[22] M. Beuthe, Towards a unique formula for neutrino oscillations in vacuum, Phys. Rev. D **66** (2002) 013003.

[23] E. Kh. Akhmedov and A. Yu. Smirnov, Paradoxes of neutrino oscillations, Phys. Atom. Nucl. **72** (2009) 1363-1381.

[24] M. Beuthe, Oscillations of neutrinos and mesons in quantum field theory, Phys. Rept. **375** (2003) 105-218.

[25] D. Kruppke, On Theories of neutrino oscillations: A Summary and characterisation of the problematic aspects, Diploma Thesis, 2007.

[26] L. B. Okun and I. S. Tsukerman, Comment on equal velocity assumption for neutrino oscillations, Mod. Phys. Lett. A **15** (2000) 1481-1482.

[27] M. C. Gonzalez-Garcia, M. Maltoni, J. Salvado and T. Schwetz, Global fit to three neutrino mixing: Critical look at present precision, JHEP **1212** (2012) 123.

[28] R. P. Feynman, The Feynman lectures on physics, Vol, 3rd, ed., Addison Wesley Longman, 1970.

[29] J. Kameda, Detailed studies of neutrino oscillations with atmospheric neutrinos of wide energy range from 100 MeV to 1000GeV in Super-Kamiokande, Doctoral Thesis, University of Tokyo, 2002

[30] A. S. Joshipura and K. M. Patel, Horizontal symmetries of leptons with a massless neutrino, Phys. Lett. B **727** (2013), 480-487.

[31] F. Benatti and R. Floreanini, Massless neutrino oscillations, Phys. Rev. D **64** (2001) 085015.

[32] S. I. Kruglov, Equations for massless and massive spin-1/2 particles with varying speed and neutrino in matter, Int. J. Mod. Phys. A **29** (2014) 1450031.

[33] J. J. Carey, J. Zawadzka, D. A. Jaroszynski and K. Wynne, Noncausal time response in frustrated total internal reflection? Phys. Rev. Lett. **84** (2000) 1431-1434.

[34] Ph. Balcou and L. Dutriaux, Dual optical tunneling times in frustrated total internal reflection, Phys. Rev. Lett. **78** (1997) 851-854.

[35] LHCb collaboration, A comprehensive analysis of the $B^0 \to K^{*0}\mu^+\mu^-$ decay, arXiv: 2512.18053.

[36] S. Navas, et al.(Particle Data Group), Phys. Rev. D 110 (2024) 030001

[37] P. -Z. Ning, L L., D. -F. Min, Fundamental nuclear physics-nucleons and nuclei, Higher Education Press, Beijing, 2003.